\documentclass[twocolumn,english,aps,nofootinbib]{revtex4}
\usepackage[T1]{fontenc}
\usepackage[latin9]{inputenc}
\usepackage{amsmath}
\usepackage{graphicx}
\usepackage{amssymb}

\makeatletter
\@ifundefined{textcolor}{}
{%
 \definecolor{BLACK}{gray}{0}
 \definecolor{WHITE}{gray}{1}
 \definecolor{RED}{rgb}{1,0,0}
 \definecolor{GREEN}{rgb}{0,1,0}
 \definecolor{BLUE}{rgb}{0,0,1}
 \definecolor{CYAN}{cmyk}{1,0,0,0}
 \definecolor{MAGENTA}{cmyk}{0,1,0,0}
 \definecolor{YELLOW}{cmyk}{0,0,1,0}
 }

\makeatother

\makeatother

\usepackage{babel}
\usepackage[unicode=true, 
 bookmarks=true,bookmarksnumbered=false,bookmarksopen=false,
 breaklinks=false,pdfborder={0 0 1},backref=false,colorlinks=false]
 {hyperref}
\hypersetup{pdftitle={Underwater bubble pinch-off: transient stretching flow},
 pdfauthor={Daniel C. Herbst, Wendy W. Zhang}}

\begin{document}

\preprint{This line only printed with preprint option}

\title{Underwater bubble pinch-off: transient stretching flow}
\thanks{{\copyright}2011 American Physical Society}

\author{Daniel C. Herbst}
\email{herbst@uchicago.edu}

\affiliation{Physics Department and the James Franck Institute, University of
Chicago, Chicago IL 60637}

\author{Wendy W. Zhang}

\affiliation{Physics Department and the James Franck Institute, University of
Chicago, Chicago IL 60637}


\begin{abstract}
At the point of pinch-off of an underwater air bubble, the speed of
water rushing in diverges. Previous studies that assumed radial flow
throughout showed that the local axial shape is two smoothly connected,
slender cones that transition very slowly (logarithmically) to a cylindrical
segment. Our simulations show that even with initially radial flow,
a transient vertical flow develops with comparable speeds. Bernoulli
pressure draws water into the singularity region while incompressibility
forces it away from the neck minimum, generating significant vertical
flows that rapidly slenderize and symmetrize the collapse region.
This transition is due to a different mechanism, occurring much faster
than previously expected. Vertical flows dictate the neck shape evolution.
\end{abstract}
\maketitle
\begin{figure}[tbh]
\begin{centering}
\includegraphics[clip]{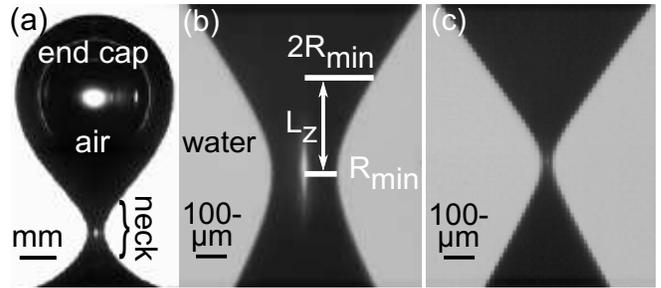}
\par\end{centering}

\caption{\label{fig:experimental_images}Pinch-off of bubble (dark region)
from nozzle while submerged in water (white). Bright spots are optical
artifacts. (a) Initially, the bubble neck shape near the minimum has
a generic quadratic profile. (b-c) Near pinch-off, the shape becomes
two cones connected at the vertex by a short segment. A characteristic
vertical length scale $L_{z}$ for the neck is the distance from the
minimum to a height where the neck radius is $2R_{\mathrm{min}}.$
(Images courtesy of N. C. Keim and S. R. Nagel).}

\end{figure}

\emph{Introduction}--Mathematical models of physical processes often
predict the formation of a singularity. Initially smoothly distributed
and finite physical quantities, such as velocity and pressure, diverge
in a finite time. Examples include supernovae, gravitational collapse
into black holes, and the pinch-off of a fluid drop. Most of the first-known
singularities exhibited self-similarity and universality. Self-similarity
indicates that the system, as it approaches the critical time $\left(t\rightarrow t^{*}\right),$
is nearly identical at different times except for some rescaling by
a function of time to singularity $\left(t^{*}-t\right),$ for example
a system radius. Universality indicates that this function is independent
of initial conditions and boundary conditions, so that every occurrence
of a given type of singularity occurs exactly the same way. An example
of a universal singularity is the pinch-off of a water drop in air,
in which surface tension smooths out azimuthal shape vibrations \cite{ting(keller)1990,Sirovich:1994:TPA:186479,shi1994,RevModPhys.69.865,Day1998}.
We focus on the opposite phenomenon, namely the pinching off of an
air bubble underwater (Fig.~\ref{fig:experimental_images}). This
commonplace phenomenon exhibits neither self-similarity nor universality,
and its low energy makes it amenable to study using high-speed photography
on a table-top setup \cite{burton05,PhysRevLett.97.144503,thoroddsen:042101}.

A nozzle submerged underwater blows a buoyant air bubble, which eventually
detaches from the nozzle. Initially, surface tension dominates and
the bubble neck shape is quadratic, with a slight asymmetry due to
the hydrostatic pressure gradient (Fig.~\ref{fig:experimental_images}a).
As the water rushes in faster, inertial forces overtake surface tension
as the dominant driver, while viscosity continues to be negligible.
The neck shape takes the form of a hyperbola of rotation \cite{longuet-higgins1991}.
As the system approaches the singularity, the two cones become more
slender. Previous theory and simulation focused on the long-and-slender
regime, where vertical flows are negligible \textit{a priori}, so
that individual vertical cross sections evolve independently~\cite{longuet-higgins1991,oguz93,PhysRevLett.95.194501,egg07,PhysRevE.80.036305}.
There, the approach to cylindrical is very slow (logarithmic in $R_{\mathrm{min}}).$
We focus on the transient regime where the vertical flows dominate
the shape evolution and occur much faster. Finally, at small length
scales, the singularity is pre-empted by one of two effects. One is
airflow in the neck, which becomes important when the neck aspect
ratio reaches $L_{z}/R_{\mathrm{min}}\sim\mathrm{\sqrt{\rho_{liquid}/\rho_{gas}}},$
or about 30 for air-water systems \cite{PhysRevLett.98.144503,PhysRevLett.104.024501}.
Second, if the neck is not perfectly axisymmetric, azimuthal vibrations
are generated \cite{PhysRevLett.97.144503,schmidt2009,Turitsyn2009}.
As pinch-off proceeds, the amplitudes of the excited vibrations remain
constant. When the neck radius shrinks down to the length scale of
the vibrations, the sides contact before the void is filled. 

These vibrations are what excludes the bubble pinch-off from being
self-similar and universal. Analogous instances of memory-encoding
vibrations arise, for example, in the implosion of shock fronts \cite{whitham57}
and implosion of spherical voids. All such systems have two commonalities:
damping forces become negligible compared to inertia, and the flow
is predominantly inward collapse. The second condition, specifically
radial flow, was assumed in previous studies of azimuthal vibrations
in bubble pinch-off~\cite{schmidt2009,Turitsyn2009}, so the behavior
could be quite different with vertical flow present. 

In addition, we are interested in whether there is memory of the axial
shape. This type of memory is present when water pinches off in a
bath of viscous oil: the axial curvature persists until pinch-off~\cite{doshi03}.
In that system, the viscosity dampens any azimuthal vibrations. Since
the flow is purely radial, however, each cross section evolves independently.
Incidentally, the collapse velocity is independent of height, and
the curvature persists. 

Previous theory and simulation has only studied air bubble pinch-off
in long-and-slender regimes, but pinch-off in other regimes is relevant.
Inviscid collapse of arbitrarily shaped voids is important, for example,
in cavitation bubbles generated by complex flows, ultrasound, or laser
pulses~\cite{Lim_Ohl2010}; in hull breach; or in an object plunging
into water \cite{bergmann09}. The collapse of these voids is also
controlled by inertia, but they are not necessarily in the long-and-slender
limit, and significant vertical flows may be present. With that in
mind, we simulate shapes with very squat and/or asymmetric cones under
generic velocity profiles. Remarkably, no matter how exaggerated the
initial conditions, strong vertical flows rapidly drive the shape
to two slender, symmetric cones (assuming axisymmetry). This happens
within observable time frames, before the singularity is pre-empted.
Therefore, every axisymmetric cavity pinches off with the well known
long-and-slender behavior.

\emph{Problem formulation}--We use the fact that the inertia of the
water inflow dominates viscosity, surface tension, and gravity in
the limit the minimum radius $R_{\mathrm{min}}\rightarrow0.$ Therefore,
we set those ignorable terms to zero. The exterior velocity field
$\mathbf{u}$ is incompressible $(\mathbf{\nabla}\cdot\mathbf{u}=0),$
irrotational $(\mathbf{\nabla}\times\mathbf{u}=\mathbf{0}),$ and
decays to zero far away. We also assume that the air in the bubble
is dynamically passive, with a uniform pressure $P_{\mathrm{air}}(t)$
whose value ensures constant bubble volume in time. 

Being curl-free, the velocity can be described by a scalar potential,
defined as $\mathbf{\nabla}\phi\equiv\mathbf{u}.$ Equating the relevant
stresses \emph{on the interface} gives a non-linear differential equation
for $\phi,$ first-order in time and space: \begin{equation}
P_{\mathrm{air}}+\rho\left(\frac{\partial\phi}{\partial t}+\frac{1}{2}\left|\mathbf{\nabla}\phi\right|^{2}\right)=0.\label{eq:stressboundary}\end{equation}

\begin{figure}
\begin{centering}
\includegraphics{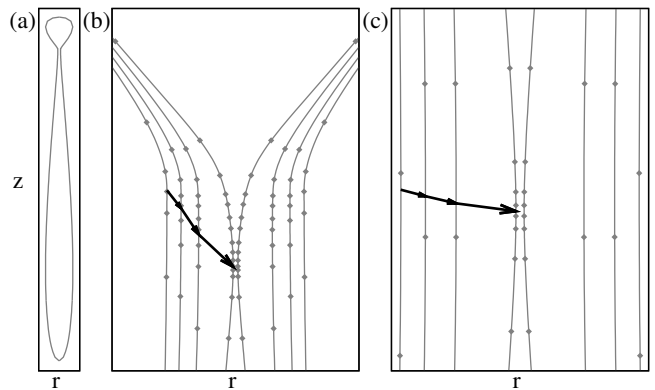}
\par\end{centering}

\caption{\label{fig:asymmetric_evolution}Pinch-off shifts and distorts the
neck shape near the minimum from a severely asymmetric hyperbola into
a symmetric one by reducing the opening angle of the larger {}``cone''
until it equals the smaller opening angle. (a) Initial surface is
comprised of a small bubble connected to a larger, elongated one.
The upper opening angle $\theta_{+}=40^{\circ},$ while the lower
opening angle $\theta_{-}=5^{\circ}.$ The initial flow is radial.
(b)-(c) successively magnify by 50x. Arrows indicate movement of the
neck minimum. (b) $\Delta t=0.19,$ and (c) $\Delta t=4.4\cdot10^{-4}.$
All times non-dimensionalized by $R_{0}/u_{0}.$ Each innermost profile
becomes the outermost in the next image. Every tenth node is shown. }

\end{figure}

The kinematic condition, which says that surface elements are advected
by$\left.\frac{\mathrm{d}\mathbf{x}}{\mathrm{d}t}\right|_{S}=\left.\mathbf{\nabla}\phi\right|_{S},$
determines how the interface evolves. Since the exterior flow is incompressible,
$\mathbf{\nabla}^{2}\phi=0,$ i.e., Laplace's equation holds. This
allows $\mathbf{u}$ to be solved everywhere in the exterior given
$\phi$ on the surface. 

We use the co-moving derivative $\frac{D\phi}{Dt}=\frac{\partial\phi}{\partial t}+\mathbf{u}\cdot\mathbf{\nabla}\phi$
in equation \ref{eq:stressboundary} to evolve $\phi$ for discrete
fluid elements. Also, we only need $u_{\perp}\equiv\mathbf{\nabla}\phi\cdot\mathbf{\hat{n}}$
on the interface, so we use the Green's integral form for $\phi$
on the surface:

\begin{equation}
2\pi\phi\left(\mathbf{x}_{0}\right)=\oint\limits _{\mathbf{x}_{1}\in S}\left(\phi\frac{\left(\mathbf{x}_{1}-\mathbf{x}_{0}\right)}{\left|\mathbf{x}_{1}-\mathbf{x}_{0}\right|^{3}}+\frac{\nabla\phi}{\left|\mathbf{x}_{1}-\mathbf{x}_{0}\right|}\right)\cdot\mathbf{\hat{n}}\mathrm{d}S,\label{eq:green}\end{equation}
 where $\mathbf{x}_{0}$ is a fixed point on $S,$ $\mathbf{x}_{1}$
parameterizes the surface, and $\mathbf{\hat{n}}$ is the unit surface
normal at $\mathbf{x}_{1}$ pointing into the bubble \cite{pozrikidis}.

\begin{figure*}[!t]
\begin{centering}
\includegraphics[clip]{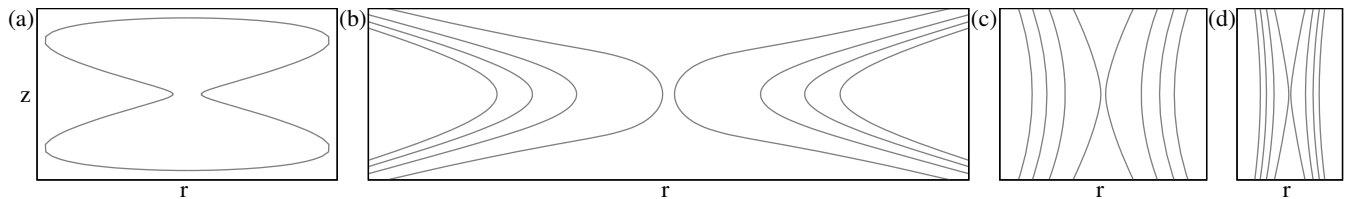}
\par\end{centering}

\caption{\label{fig:symmetric_evolution}Pinch-off stretches a bubble neck
that initially is symmetric about the minimum toward a cylindrical
segment. (a) Hyperbola with opening angle $\theta=75^{\circ}$ and
radial flow. (b)-(d) successively magnify by 12x. Time between snapshots:
(b) $\Delta t=0.28,$ (c) $\Delta t=2.4\cdot10^{-3},$ and (d) $\Delta t=7.4\cdot10^{-6}.$
Each innermost profile becomes the outermost in the next image. }

\end{figure*}

With both governing equations defined \emph{on the interface}, the
formulation reduces to 2 dimensions. We further reduce to 1 dimension
by assuming axisymmetry. At each time step, we use cubic splines to
interpolate the $N(t)$ discrete surface nodes $\mathbf{x}_{i}$.
Then, for each $i\in\left\{ 1\ldots N\right\} $ we allow $\mathbf{x}_{i}$
to take the place of $\mathbf{x}_{0}$ in equation \ref{eq:green}.
We use Gaussian quadrature to perform the integral over segments connecting
adjacent spline midpoints, with the first and last segments being
only half splines. This gives the needed relationship for $\left\{ u_{i}^{\perp}\right\} $
in terms of $\left\{ \phi_{i}\right\} $. Using $\delta t\sim R_{\mathrm{min}}/\dot{R}_{\mathrm{min}},$
we advect the $N$ nodes using $\mathbf{u}$ and evolve $\phi$ using
equation \ref{eq:stressboundary}, completing the cycle.

In order to accurately resolve the pinch-off dynamics, we found it
important to use a node distribution scheme that maintains a gradual
variation in the spacing between node points and that continually
adds nodes in the neighborhood of the minimum (see Fig. \ref{fig:asymmetric_evolution}).
At each moment, if the vertical distance $\Delta z$ between nodes
exceeds a maximum spacing $\Delta z_{\mathrm{max}}(t,\, z)$ in the
region where the neck radius is less than $4R_{\mathrm{min}},$ we
add a new node at the midpoint of that spline. After experimenting
with several functions for $\Delta z_{\mathrm{max}},$ we found $\Delta z_{\mathrm{max}}=\max\left[0.15\min\left(\left|z\right|,L_{z}\right),0.08L_{z}\right]$
allows the simulation to accurately track the dynamics %
\footnote{For the up-down asymmetric case, we define the current height of the
neck minimum as $z_{0},$ $R\left(z_{0}\pm L_{z}^{\pm}\right)=2R_{\mathrm{min}},$
and $\Delta z_{\mathrm{max}}=\max\left[0.15\min\left(\left|z-z_{0}\right|,L_{z}^{+}\right),0.08\min\left(L_{z}^{+},L_{z}^{-}\right)\right]$
for $z\geq z_{0}$ (reverse superscripts for $z<z_{0}).$ %
}. 

We begin the simulation by specifying an interface shape and velocity
field. In general, there are an infinite number of each. Here, we
examine generic smooth distributions. To expedite the computation,
we do not prescribe a purely quadratic shape profile at $t=0,$ but
instead use the result from previous studies that a slender quadratic
neck evolves into a hyperbola, and prescribe:

\[
R(z)=R_{0}\sqrt{1+\tan^{2}\theta\left(\frac{z}{R_{0}}\right)^{2}\left(1-\frac{\left(\tan^{2}\theta\right)\left(z^{2}\right)}{4\left(R_{\mathrm{e}}^{2}-R_{0}^{2}\right)}\right)},\]
where $2\theta$ is the opening angle of the cone, $R_{0}\equiv R_{\mathrm{min}}\left(t=0\right),$
and $R_{\mathrm{e}}=10R_{0}$ is the radius of the end cap. We experimented
with different smoothly-varying initial velocity fields, with corrections
in the end caps to preserve bubble volume. One such field is given
by the normal velocity on the surface: $u_{\perp}=u_{0}R_{0}\left[R_{0}^{2}+\tan^{2}\xi\left(1+\tan^{2}\xi\right)z^{2}\right]^{-1/2}-\gamma s^{4}.$
Incompressibility then determines the tangential component of the
velocity. The $\gamma s^{4}$ is a correction term necessary for bubble
volume conservation, where $s$ is the arc length along the surface
from the neck minimum and $\gamma$ is a constant. The parameter $\xi$
specifies the initial orientation of the vertical velocity in the
neck region. If $\xi>\theta,$ the vertical velocities are directed
toward the midline, compressing the neck aspect ratio into a more
squat shape. This configuration, though, is unstable. Very soon $(R_{\mathrm{min}}\gtrsim0.1R_{0})$
the vertical velocities near the midline flip to orient away from
the midline and proceed to stretch the neck. If, on the other hand,
the system is initialized with $\xi<\theta,$ the vertical velocities
point away from the midline and continue to point away, stretching
the neck for the entire collapse. After their respective transients,
however, both situations fall into the same dynamics, only differing
by a time offset. Therefore, for the remainder of the paper we consider
an initial radial flow, which corresponds to $\xi=\theta.$ To specify
a radial flow for up-down asymmetric shapes, we use an explicit radial
flow with end-cap corrections: $u_{\perp}=\frac{-u_{0}R_{0}}{r}\mathbf{\hat{\mathbf{r}}}\cdot\mathbf{\hat{n}}-\gamma s^{4}.$

\begin{figure}

\includegraphics{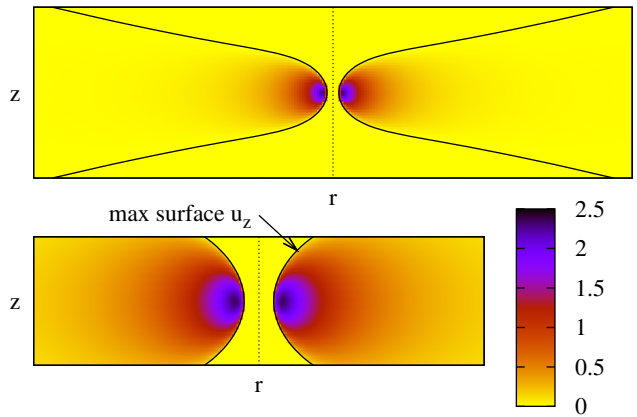}\caption{\label{fig:pressure}(Color online) A toroidal pressure peak develops
just inside the neck minimum that has a cross-section aspect ratio
roughly $L_{z}$ by $R_{\mathrm{min}}$ and tracks the neck inward
with speed $\dot{R}_{\mathrm{min}}.$ Water is accelerated away from
the peak, generating vertical velocity on the surface comparable in
magnitude to radial velocity. The point of max $u_{z}$ on the surface
is indicated. This pressure distribution is robust to changes in initial
shape and velocity and develops quickly. The case shown is $R_{min}/R_{0}=0.034$
for an initial shape with $\theta=75^{\circ}$ and radial flow. (Top)
corresponds to Fig. \ref{fig:symmetric_evolution} (b), innermost
profile. (Bottom) zoomed 5x. Pressure is in units $\rho u_{0}^{2}.$ }
 
\end{figure}

\emph{Results}--Small values of $\theta$ with radial flow correspond
to the regime of the slender-body approximation employed in previous
works. Our results (not shown here) agree quantitatively with their
results in this regime. Here, we focus on large values of $\theta.$
Since the bubble break-up experiment shows a slight up-down asymmetry,
we first examine a severely up-down asymmetric shape. We produce such
an initial state (Fig. 2a) by using different values of $\theta$
for the top and bottom portions of the bubble instead of prescribing
a hydrostatic pressure gradient. The upper portion of the bubble is
chosen to have the larger opening angle. We set the initial flow to
be radial in the neck region. Figs. \ref{fig:asymmetric_evolution}b-c
show the shape evolution. Initially, the minimum shifts toward the
smaller cone. As the minimum radius decreases, the upper portion of
the bubble surface rolls while the lower portion more or less preserves
the same profile. As a consequence, the profile near the minimum quickly
approaches a symmetric shape. Simulations starting with various choices
for up-down asymmetry yield the same qualitative outcome (not shown).
We have found that an up-down asymmetric initial state always proceeds
through a 3-stage evolution. First, it evolves into a symmetric profile
as described. Second, the symmetric cones become slender near the
pinch-off point. Finally, there is a slow transition to a cylinder.
Hereafter we will focus on the second stage, which connects the asymmetry
in experiment to prior works focusing on the last stage \cite{PhysRevLett.95.194501,egg07,PhysRevE.80.036305}. 

\begin{figure}[tbh]
\begin{centering}
\includegraphics[clip]{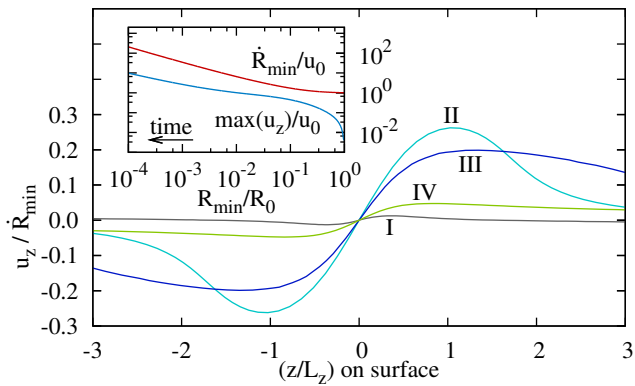}
\par\end{centering}

\caption{\label{fig:velocity_profiles}(Color online). (Inset) Both the maximum
vertical velocity on the surface, $u_{z},$ and the radial collapse
speed $\dot{R}_{\mathrm{min}}$ diverge as pinch-off approaches. (Main
figure) Dividing the vertical velocity profiles from different times
by $\dot{R}_{\mathrm{min}}$ and $z$ by $L_{z}$ scale out most of
the variation, showing that the velocity field is controlled by the
break-up dynamics. The rescaled profiles evolve non-monotonically.
The vertical velocity ratio at early times (I, $R_{\mathrm{min}}=0.9R_{0}$)
reaches a maximum at $R_{\mathrm{min}}=0.075R_{0},$ (II), then broadens
in space (III). Eventually it slowly vanishes in the last stage, which
is dominated by radial flow (IV, $R_{\mathrm{min}}=10^{-4}R_{0}$).}

\end{figure}

Fig. \ref{fig:symmetric_evolution} displays the evolution of a symmetric,
large initial opening-angle shape, $\theta=75^{\circ},$ initialized
with radial flow. Fig. \ref{fig:symmetric_evolution}b shows that
the shape quickly evolves into a nearly-cylindrical segment connecting
two cones. As the pinch-off proceeds and we zoom in close to the minimum,
the shape becomes long and slender. The effective half opening-angle
of the innermost profile in each sub-figure is (a) $\theta_{\mathrm{eff}}=75.0^{\circ},$
(b) $\theta_{\mathrm{eff}}=32.3^{\circ},$ (c) $\theta_{\mathrm{eff}}=13.3^{\circ},$
and (d) $\theta_{\mathrm{eff}}=7.6^{\circ}$ %
\footnote{$\theta_{\mathrm{eff}}$ defined by $2R_{\mathrm{min}}=R_{\mathrm{min}}\sqrt{1+\left(\tan^{2}\theta_{\mathrm{eff}}\right)\left(L_{z}/R_{\mathrm{min}}\right)^{2}}$%
}. This is a very fast process compared to the final stage, which roughly
begins where (d) leaves off, where the shape very slowly approaches
a cylinder.  Fig. \ref{fig:pressure} shows why this takes place.
Immediately after starting, the water develops a pressure peak just
inside the neck. In order to satisfy the boundary condition $P=P_{\mathrm{air}}$
on the surface, the pressure peak takes on an aspect ratio similar
to the neck aspect ratio. The pressure gradients generate vertical
velocity on the surface comparable to $\dot{R}_{\mathrm{min}}.$ Although
$u_{z}$ is zero at the neck minimum due to the symmetry, it becomes
large just a short distance along the surface. Fig. \ref{fig:velocity_profiles}
shows the surface velocity fields. The inset shows that the vertical
velocity diverges approximately in pace with the radial velocity.
Therefore, we consider the rescaled vertical flow, $u_{z}(z,t)/\dot{R}_{\mathrm{min}},$
in the neighborhood of the minimum (Fig. \ref{fig:velocity_profiles},
versus $z$ rescaled by $L_{z}).$ The rescaling shows that $u_{z}$
always remains a small fraction of the radial collapse, reaching a
maximum in the transient period (II) and then broadening in spatial
extent.

\begin{figure}[!tbh]
\begin{centering}
\includegraphics[clip]{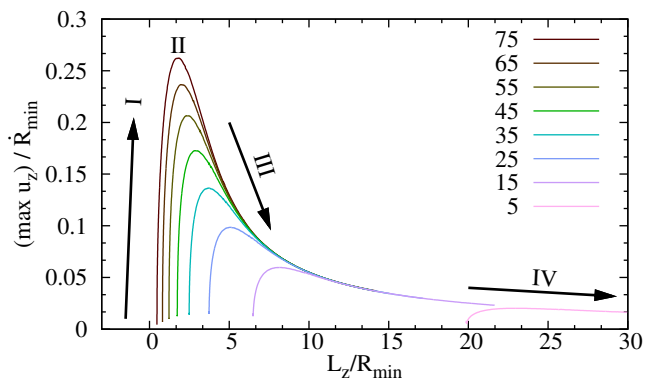}
\par\end{centering}

\caption{\label{fig:symmetric_phase_diagram}(Color online) Normalized maximum
vertical velocity, $\max\left|u_{z}\right|/\dot{R}_{\mathrm{min}},$
versus shape aspect ratio, for symmetric shapes with initially radial
flow, labeled by $\theta$ in degrees (from top to bottom). The flow
invariably becomes a stretching flow, increasing the aspect ratio
of the break-up region, $L_{z}/R_{\mathrm{min}}.$ The normalized
maximum velocity initially increases rapidly (I), peaks (II) and finally
decays onto one curve (III, IV). }

\end{figure}

We show the continuous evolution of the velocity distributions for
$\theta=75^{\circ}$ by plotting the maximum of $u_{z}/\dot{R}_{\mathrm{min}}$
versus the aspect ratio (Fig. \ref{fig:symmetric_phase_diagram}a,
uppermost curve). Roman numerals indicate corresponding stages between
Fig. \ref{fig:velocity_profiles} and Fig. \ref{fig:symmetric_phase_diagram}a.
Just after $t=0,$ a stretching flow builds in strength before the
shape has time to react (I). After reaching a peak vertical-to-radial
ratio (II), the neck continues to stretch (III). The exterior flow
decays slowly to radial implosion. Starting with different initial
opening angles yields the same qualitative behavior. Smaller initial
opening angles generate smaller maximum $u_{z}/\dot{R}_{\mathrm{min}}$
in the transient regime (I-II). Eventually all initial conditions
enter the slender-body regime (IV) where we find the slow evolution
towards a perfect cylinder described in previous works \cite{PhysRevLett.95.194501,egg07,PhysRevE.80.036305}. 

\emph{Discussion}--While our analysis has focused on axisymmetric
dynamics, the conclusions should remain relevant when the neck shape
has a slight azimuthal asymmetry since prior linear stability studies
have shown that the asymmetry does not grow rapidly but instead persists
as vibrations of fixed amplitude. In the opposite regime when the
neck shape is strongly distorted from axisymmetry, experiments show
the bubble neck rips itself apart in a series of violent, apparently
discrete jerks \cite{PhysRevLett.97.144503}. The force balance controlling
this break-up mode remains an open question.

\emph{Conclusions}--We showed here that strong, transient vertical
flows are the dominant effect in the transition of the neck shape
from large to small cone angles. This transition happens as a distinct,
transient phase, as opposed to the very slow, continuous transition
predicted by previous studies assuming weak vertical coupling. The
transition is effectively complete once the minimum radius has decreased
by a factor of $100,$ an easily observable window. Even if the initial
cone angles are very large, the system just induces a stronger vertical
flow, still guaranteeing a rapid transition. Moreover, if the vertical
velocity is initially compressing, it quickly reorients to a strong
stretching flow. All these things taken together indicate that the
transition to small cone angles will occur before any cut-off length
scale. After the magnitude of the vertical flow peaks relative to
the radial flow, it soon becomes small in comparison. This guarantees
that the system will reach the slender-body phase characterized by
a logarithmically slow transition to cylindrical \cite{PhysRevLett.95.194501,egg07,PhysRevE.80.036305},
along with memory of azimuthal vibrations \cite{PhysRevLett.97.144503,schmidt2009,Turitsyn2009},
both studied extensively. In sum, the radial-flow dominated implosion
singularity controls the final stage of the dynamics even in situations
with strong vertical coupling.
\begin{acknowledgments}
This work was supported by NSF No.~CBET-0967282, NDSEG fellowship
(DCH) and the Keck initiative for ultrafast imaging (University of
Chicago). Monte Rinebold experimented with parameters to evolve up-down
asymmetric shapes. We thank Justin Burton, Nathan Keim, Lipeng Lai,
Sidney Nagel, and Laura Schmidt for discussions and encouragement.
We also thank the anonymous referees for helpful input.
\end{acknowledgments}
\bibliographystyle{apsrev}

\end{document}